\documentclass[aps,prb,twocolumn,groupedaddress,showpacs,amsmath]{revtex4}

\usepackage{graphicx,amssymb}

\begin{document}

%

\title{Field-induced level crossings in spin clusters: Thermodynamics and magneto-elastic instability}

\author{O. Waldmann}
 \email[E-mail: ]{waldmann@iac.unibe.ch}
 \affiliation{Department of Chemistry and Biochemistry, University of Bern, CH-3012 Bern, Switzerland}

\date{\today}

\begin{abstract}
Quantum spin clusters with dominant antiferromagnetic Heisenberg exchange interactions typically exhibit a
sequence of field-induced level crossings in the ground state as function of magnetic field. For fields near
a level crossing, the cluster can be approximated by a two-level Hamiltonian at low temperatures.
Perturbations, such as magnetic anisotropy or spin-phonon coupling, sensitively affect the behavior at the
level-crossing points. The general two-level Hamiltonian of the spin system is derived in first-order
perturbation theory, and the thermodynamic functions magnetization, magnetic torque, and magnetic specific
heat are calculated. Then a magneto-elastic coupling is introduced and the effective two-level Hamilitonian
for the spin-lattice system derived in the adiabatic approximation of the phonons. At the level crossings the
system becomes unconditionally unstable against lattice distortions due to the effects of magnetic
anisotropy. The resultant magneto-elastic instabilities at the level crossings are discussed, as well as the
magnetic behavior.
\end{abstract}


\maketitle

%

\section{Introduction and conclusions}

The issue of level crossings (LCs) in quantum spin clusters is of broad interest. The most elementary example
is probably that of an effective spin-1/2 with an avoided LC in a magnetic field, where the tunneling
probability at constant field-sweep rate is described by the Landau-Zener-Stueckelberg model.\cite{LZS}
Experimental realizations are, for instance, the single-molecule magnets Mn$_{12}$ or Fe$_8$.\cite{Mn12_Fe8}

In antiferromagnetic (AFM) spin clusters, LCs also emerge from the interplay of Heisenberg exchange
interactions in the cluster and the Zeeman term, which leads to a sequence of LCs in the ground state as
function of magnetic field.\cite{Taf94,Gat94,Cor99a,Cor99b,OW_XFe6,OW_CsFe8,Sha02} More precisely, spin
clusters are considered which are modeled by the microscopic spin Hamiltonian
\begin{equation}
\label{eqn:Hm}
  \hat{H} = -\sum^N_{i \neq j}{ J_{ij} \hat{\textbf{S}}_i \cdot \hat{\textbf{S}}_{j} } +
 \mu_B g \hat{\textbf{S}} \cdot \textbf{B} + \hat{H}_1
\end{equation}
consisting of the Heisenberg exchange term and the Zeeman term ($N$ is the number of spin centers in the
cluster and $\hat{\textbf{S}} = \sum^N_i \hat{\textbf{S}}_i$ is the total spin operator). The term
$\hat{H}_1$ contains all remaining relevant terms, such as magnetic anisotropy (single-ion anisotropy,
anisotropic exchange, dipole-dipole interactions, etc.) or couplings to the environment (phonons,
intercluster magnetic couplings, nuclear spins, etc.), but should be small in the sense of a perturbation.
Then the eigenstates of $\hat{H}$ can be classified by the total spin quantum numbers $S$ and $M$. The AFM
Heisenberg interactions typically lead to an energy level scheme, where the energy of the lowest state for
each value of $S > S_0$ increases according to $E(S_0) < E(S_0+1) < \ldots < E(S_{max})$ ($S_0$ is the
ground-state spin, and $S_{max} = \sum^N_i S_i$). As function of field, the Zeeman splitting produces a
sequence of LCs, at which the ground state changes from $|S=S_0,M=-S_0\rangle$ to $|S=S_0+1,M=-S_0-1\rangle$,
to $|S=S_0+2,M=-S_0-2\rangle$, and so on.\cite{Sha02} For magnetic fields close to the LC conditions, the
system becomes almost degenerate, and is well described as a two-level system at low temperatures.

Studies on this type of LCs are interesting from several perspectives, and numerous experimental examples are
known both in the area of molecular nanomagnets and inorganic magnetic compounds. For instance, the
determination of the LC fields from field-dependent measurement of thermodynamic quantities, such as the
magnetization or magnetic torque, directly yields the energies $E(S_0+1)-E(S_0)$, $E(S_0+2)-E(S_0)$, etc., of
the next higher spin levels. This "thermodynamic" spectroscopy often enables a precise determination of the
exchange coupling constants in the cluster.\cite{Taf94,Gat94,Cor99a,Cor99b,OW_XFe6,OW_CsFe8,Sha02} In some
molecular nanomagnets, such as the AFM wheels, the energy of the lowest spin levels increases quadratically
with $S$ according to $E(S) \propto S(S+1)$,\cite{Taf94,Sla02,Sch00,OW_SPINDYN} which reflects the quantized
rotation of the Ne\'el vector in these clusters.\cite{And52,Ber92,OW_Cr8,OW_CCR}

Interesting phenomena also emerge from the fact, that near a LC the system is rather susceptible to the
perturbations represented by $\hat{H}_1$. Among the molecular nanomagnets, the AFM
wheels\cite{Taf94,AFwheels} and related systems, such as the Mn-[3~$\times$~3] grid or the modified
wheels,\cite{OW_Mn3x3,OW_QMO,Mei01,Lar03} have become prototypical examples in this regard. The most
important term in $\hat{H}_1$ usually is the single-ion anisotropy. In the wheels, however, because of their
nominally high symmetry, the LCs would remain true LCs despite this perturbation,\cite{OW_FW_QT,Aff02} but
with small deviations from ideal cyclic symmetry spin terms such as a Dzyaloshinsky-Moriya interaction may
mix the two levels at a LC, thus leading to avoided LCs.\cite{Aff02,Cin02} Indications of this has been
reported for some ferric wheels.\cite{Aff02,OW_NaFe6,Cin02} In \emph{a priory} less symmetric molecules the
symmetry restrictions are absent, and the single-ion anisotropy becomes efficient in mixing the spin levels
at the LCs.\cite{Car03b,OW_QMO,OW_NVT_3x3,Car05} The result can be dramatic. For instance, the magnetic
torque at low temperatures may display a pronounced oscillatory field dependence, i.e., show quantum
magneto-oscillations. This phenomenon has been observed in the Mn-[3~$\times$~3] grid,\cite{OW_QMO} and two
modified wheels.\cite{Car05} Since levels with different values of $S$ are mixed, the total spin
$\hat{\textbf{S}}$ exhibits quantum fluctuations at the LCs, which could enable coherent oscillations of the
total spin if dissipation is weak.\cite{Car03b} Moreover, the torque oscillations observed in
Mn-[3~$\times$~3] provided the experimental demonstration of the so-called tower of states in a finite AFM
square lattice.\cite{OW_CCR}

$\hat{H}_1$ also may include perturbations such as a coupling to the nuclear spins or the phonon system. In
fact, the LCs in the molecular wheels were intensively studied by nuclear magnetic resonance
(NMR).\cite{Jul99,Aff02,Bor04,Pil05,Mic05} Also effects of the spin-phonon coupling were observed in the
wheels NaFe$_6$ and Fe$_{12}$,\cite{OW_NaFe6,Ina03} and other clusters,\cite{Sha99} where it resulted in
magnetic butterfly-hysteresis at the LCs due to a phonon-bottleneck effect. From a general perspective, at
the LCs the clusters represent two-level systems with dissipation.\cite{Leg87} They open attractive
experimental opportunities for exploring this model, since, for instance, the strength of the mixing at the
LCs and hence the influence of dissipation may be tuned by external parameters such as the orientation of the
magnetic field.\cite{OW_QMO,OW_NVT_3x3,Car05}

Other interesting examples are found in inorganic compounds, for instance, in the materials characterized as
weakly-interaction dimers. Prototypical systems are TlCuCl$_3$ and BaCuSi$_2$O$_6$.\cite{Nik00,Jai04} The
dominant AFM interactions in the dimers lead to a $S=0$ dimer ground state and a higher lying $S=1$ level,
which is split in a magnetic field resulting in a LC, exactly as described before. Weaker magnetic exchange
interactions between the dimers, however, drastically modify the behavior of the system for fields close to
the LC; they give rise to various field-induced magnetic long-range ordering phenomena, such as the
Bose-Einstein condensation of triplets and dimensional reduction at quantum critical
points.\cite{Aff90,Mat02,Seb06}

The (near) degeneracy of the levels at the LCs suggests also the possibility of spontaneous structural
distortions of the cluster at the LC fields, driven by the interaction of the spin system with the phonons.
In fact, in view of the resemblance of an AFM molecular wheel with an 1D AFM quantum spin chain, a
spin-Peierls type of effect seems to be obvious. However, general arguments imply that for finite spin
clusters, which exhibit a gap in their excitation spectrum, the energies should vary quadratically with the
distortion coordinate. Hence, a spontaneous structural instability would be conditional, and not likely to
occur in real materials. Furthermore, a recent theoretical study on AFM Heisenberg rings demonstrated that,
for $S_i > 1/2$, a spontaneous distortion would be of first order and so strong, that it would disrupt the
molecule.\cite{Spa04}

Recently it has been shown, however, that the magnetic anisotropy terms in $\hat{H}_1$ in fact give rise to
an unconditional instability of the clusters against spontaneous distortions at the LCs.\cite{OW_FIMEI} The
argument is based on a phenomenological model, which describes the spin system by a two-level Hamiltonian and
treats the phonons in the adiabatic limit (the effective Hamiltonian of the coupled spin-phonon system will
be called $\hat{H}_{SP}$). In the molecular wheel CsFe$_8$, phase transitions at the LC points were recently
observed.\cite{OW_FIMEI} Since the crystal structure as well as other arguments disapproved the presence of
sufficient magnetic interactions between the clusters, the findings were interpreted as magneto-elastic
instabilities at the LCs. The experimental data in fact were consistent with the predictions of the
phenomenological model $\hat{H}_{SP}$.

The primary goal of this work is to present a coherent derivation of $\hat{H}_{SP}$ and the field-induced
magneto-elastic instabilities, and to discus the resultant magnetic behavior (extending a previous brief
report\cite{OW_FIMEI}). As a preparatory step, the two-level Hamiltonian of the spin system is derived
(Secs.~II and III), and the thermodynamic magnetization, magnetic torque, and magnetic specific heat
calculated (Sec.~IV). Then the magneto-elastic coupling is introduced and the two-level Hamiltonian
$\hat{H}_{SP}$ established (Sec.~V), and the magneto-elastic instabilities discussed (Sec.~VI). The subject
of field-induced LCs has been investigated intensely in the last years by several authors. Accordingly,
several pieces of the results were obtained before.

The results should be useful in several regards. The thermodynamic results provide insight into the different
magnetic behavior of, e.g., AFM wheels and the Mn-[3~$\times$~3] grid (in the wheels, the sequence of LCs
leads to staircase-like magnetization and torque curves; in the grid, it leads to oscillations in the
torque). Furthermore, they provide a handy method to parameterize experimental data, and to extract from them
key parameters, such as the level-mixing strength. The presented model of the magneto-elastic instability has
all the drawbacks and virtues of a phenomenological model. It cannot be expected to provide a full
quantitative description of a specific material (e.g., of CsFe$_8$), or to answer specific microscopic
details, such as which distortion mode is involved. However, it is the most simple model which describes the
basic mechanism (it actually represents the mean-field model). As it is often with such models, they are very
useful for a qualitative understanding of the phenomenon, and discussion of observed behavior.

A connection between the weakly-interacting dimer compounds and AFM molecular wheels is finally noted. The
low-lying excitations in AFM wheels are well described by a dimer model, where the two spins correspond to
the total spins on each AFM sublattice.\cite{And52,OW_SPINDYN,OW_Cr8} Hence, with magnetic interactions
between the clusters, AFM wheels would perfectly mimic a weakly-interacting dimer system, and show the very
same field-induced phenomena. In the AFM wheels, however, the magnetic intercluster interactions are usually
negligible, which suggested the magneto-elastic origin of the observed field-induced phase transitions in
CsFe$_8$.\cite{OW_FIMEI} In the weakly-interacting dimer compounds the observed field-induced phase
transitions are undoubtly of magnetic origin, but also clear indications of magneto-elastic couplings have
been reported.\cite{Vya04,Cho05,Mat06} In particular, NMR experiments showed that the field-induced magnetic
ordering in TlCuCl$_3$ is accompanied by a simultaneous lattice deformation.\cite{Vya04} These findings
indicate that both intercluster magnetic interactions and magneto-elastic couplings are present in general in
a specific material, and that it only depends on the relative strength whether the one or the other is
considered as the driving force for the phase transitions at the LCs. Hence, as a conclusion, the phenomena
of the field-induced magneto-elastic instabilities and field-induced magnetic ordering appear as the two
extreme sides of one and the same medal $-$ and it might be hard to distinguish them from each other in a
specific material.

%

\section{Basics}

In this section, some general results for an effective two-level Hamiltonian (TLH) are shortly reviewed. In
matrix form the TLH is written as
\begin{equation}
\label{eqn:TLH}
 \hat{H}_{TL} = \left(\begin{array}{cc} \varepsilon_1 & \Delta/2 \\ \Delta/2 & \varepsilon_2 \end{array}
 \right),
\end{equation}
where $\varepsilon_1$ and $\varepsilon_2$ are the bare energies of the two involved levels, and $\Delta$
describes a mixing between these levels. The approximation of the TLH works well then both the gap between
the two levels and the temperature are much smaller than the gaps to the next higher lying levels in the
energy spectrum. The energies $E_{\pm}$ of the two eigenstates of the TLH are
\begin{equation}
\label{eqn:E}
 E_{\pm} = \frac{1}{2}( \varepsilon_2 + \varepsilon_1 ) \pm \frac{1}{2} \sqrt{ (\varepsilon_2 -
\varepsilon_1)^2 + \Delta^2 } \equiv E_1 \pm E_2,
\end{equation}
where $E_{+}$ ($E_{-}$) denotes the energetically upper (lower) branch of the avoided LC. The partition
function $Z = 2 \exp(-\beta E_1)\cosh(\beta E_2)$ and the free energy $F = -k_B T \ln(Z)$ are functions of
the temperature $T$, and further variables, which are abbreviated symbolically by $x$, i.e., $Z \equiv
Z(T,x)$ and $F \equiv F(T,x)$. $k_B$ is the Boltzmann constant and $\beta = 1/(k_BT)$. For some relevant
derivatives one obtains
\begin{subequations}
\begin{eqnarray}
\label{eqn:Fsa}
 \frac{\partial F(T,x)}{\partial x}  &=&
 \frac{\partial E_1}{\partial x} - \tanh(\beta E_2) \frac{\partial E_2}{\partial x}
 \cr
 &=& \frac{\partial E_-}{\partial x} +
 [1-\tanh(\beta E_2)] \frac{\partial E_2}{\partial x},
 \\\label{eqn:Fsb}
 \frac{\partial^2 F(T,x)}{\partial T^2}  &=& - k_B^2 \beta^3 [1-\tanh^2(\beta E_2)] (E_2)^2.
\end{eqnarray}
\end{subequations}
A formulation of the expressions in terms of a factor $[1-\tanh^n(\beta E_2)]$ (with integer $n$) is
convenient for a discussion of the temperature dependence: For $T \rightarrow 0$ one finds $[1-\tanh^n(\beta
E_2)] \rightarrow 0$, and in the high-temperature limit $T \rightarrow \infty$ one finds $[1-\tanh^n(\beta
E_2)] \rightarrow 1$. $E_2$ is half of the energy gap between the two levels, $E_+ - E_- = 2E_2$.

\section{Two-level Hamiltonian}

In this section, the effective TLH, which describes the behavior of a (rigid) spin cluster at a field-induced
LC, is derived in first-order perturbation theory. The spin Hamiltonian $\hat{H}$ of the system is given by
Eq.~(\ref{eqn:Hm}). The type of LCs considered in this work emerge from dominant Heisenberg interactions and
the Zeeman term in the cluster, such that the effects of the term $\hat{H}_1$ can be treated
perturbationally. For simplicity it is further assumed that the cluster anisotropy is uniaxial. Then the
magnetic field is completely specified by its magnitude $B$ and its angle $\varphi$ with respect to the
uniaxial $z$ axis. No further restrictions are imposed on the treatment in this and the next section. It is
noted that for most AFM wheels, and CsFe$_8$ in particular, the anisotropy is very well characterized as
uniaxial.\cite{Cor99a,Cor99b,OW_XFe6,OW_CsFe8,Sla02,Cin02,Pil05,OW_CsFe8_NVT_INS2,Aff03,Pil03}

The most important term in $\hat{H}_1$ is usually a single-ion anisotropy term,
\begin{equation}
\hat{H}_D = \sum_i \hat{\mathbf{S}}_i \cdot \mathbf{D}_i \cdot \hat{\mathbf{S}}_i,
\end{equation}
but also a Dzyaloshinsky-Moriya term,
\begin{equation}
\hat{H}_{DM} = \sum_{i \neq j} \mathbf{d}_{ij} \cdot (\hat{\mathbf{S}}_i \times \hat{\mathbf{S}}_{j}),
\end{equation}
or a spin-phonon coupling term, etc., may be included, depending on the situation under consideration. Since
the single-ion anisotropy is of general importance, it is convenient to introduce a $\hat{H}'_1$ via
$\hat{H}_1 = \hat{H}_D + \hat{H}'_1$. In the following and Sec.~IV, $\hat{H}'_1$ is mostly disregarded, but
it will become crucial in Secs.~V and VI.

Since the main part of the microscopic Hamiltonian $\hat{H}$ is isotropic, the appropriate zero-order states
are eigenstates of the total spin operator $\hat{\textbf{S}}$, and classified by the quantum numbers $S$ and
$M$. The two states involved in the LCs have spin quantum numbers $S$ differing by one unit and $M =
-S$.\cite{SM} The wavefunctions may be written as $|\gamma,S,-S\rangle$ and $|\gamma',S+1,-S-1\rangle$, where
$\gamma$ denotes additional quantum numbers. The obvious shorthand notation $|S\rangle$ and $|S+1\rangle$
will be used in the following. In first-order the TLH Eq.~(\ref{eqn:TLH}) is obtained with $\varepsilon_1 =
\varepsilon_S$, $\varepsilon_2 = \varepsilon_{S+1}$, and
\begin{subequations}
\begin{eqnarray}
  \varepsilon_S &=& \langle S|\hat{H}|S\rangle,\\
  \Delta/2 &=& \langle S|\hat{H}|S+1\rangle.
\end{eqnarray}
\end{subequations}
It is important to note that the quantization axis for the magnetic quantum number $M$ is along the direction
of the magnetic field, and not for instance along the uniaxial $z$
axis.\cite{Pak73,Cor99a,Cor01,OW_NVT_3x3,Efr06} For the calculation of the matrix elements, an operator such
as $\hat{S}_{iz}^2$, which is expressed in a reference frame with the quantization axis along the $z$ axis,
thus has first to be rotated to a frame with the $z'$ axis (the "new" $z$ axis) along the magnetic field.

For the diagonal matrix elements one directly calculates $\varepsilon_S = \Delta_S^0 - b S + \langle
S|\hat{H}_1|S\rangle$,\cite{Cor99a,Cor01} where $\Delta_S^0$ is the energy of the level $|S\rangle$ in zero
magnetic field and without $\hat{H}_1$, and the abbreviation $b = g \mu_B B$ is introduced. The first-order
approximation thus results in a linear field-dependence of the bare levels near a LC. The non-diagonal matrix
element reduces to $\Delta/2 = \langle S|\hat{H}_1|S+1\rangle$, since the Heisenberg and Zeman terms do not
mix states with different $S$. As an important result it follows that, in first-order, $\Delta$ does not
depend on the magnetic field, but only on the angle $\varphi$, i.e., $\Delta \equiv \Delta(\varphi)$.

Because of the importance of the single-ion anisotropy, the situation with $\hat{H}_1 = \hat{H}_D$ is
discussed in some detail. The matrix elements of $\hat{H}_D$ were calculated
previously.\cite{Cor99a,Cor99b,Cor01,Cin02,OW_NVT_3x3,Boc99} For the diagonal elements holds $\langle
S|\hat{H}_D|S\rangle = D_S S (S-1/2)(\cos^2\varphi - 1/3)$, where $D_S$ is the zero-field-splitting parameter
of the spin multiplet to which $|S\rangle$ belongs. One thus obtains
\begin{eqnarray}
\label{eqn:eps}
 \varepsilon_S(b,\varphi) &=& \Delta_S(\varphi) - b S,
\end{eqnarray}
where $\Delta_S(\varphi)$ denotes the zero-field energy of the level $|S\rangle$ due to the combined effects
of the Heisenberg interactions and the single-ion anisotropy $\hat{H}_D$. It is interesting to note that the
dependence on the angle $\varphi$ comes entirely from the zero-field gaps $\Delta_S(\varphi)$.

The magnetic field $b_0$, at which the LC occurs, is determined by the condition $\varepsilon_1(b_0) =
\varepsilon_2(b_0)$, yielding $b_0(\varphi) = \Delta_2(\varphi) - \Delta_1(\varphi)$ or
\begin{equation}
\label{eqn:b0}
  b_0(\varphi) = a + b( \cos^2\varphi - 1/3 ),
\end{equation}
with the constants $a = \Delta_2^0 - \Delta_1^0$ and $b = D_{S+1}(S+1)(S+1/2) - D_S S
(S-1/2)$.\cite{Cor99a,Cor01} The LC field $b_0$ in general exhibits an angle dependence due to the
anisotropic terms in $\hat{H}_1$. For $\hat{H}_D$ one finds the generic behavior $b_0(\varphi) \propto
(\cos^2\varphi - 1/3) + const$. This angle dependence of the LC fields has been observed in high-field torque
experiments on several molecular AFM wheels, and was used to experimentally determine the
zero-field-splitting parameters $D_S$.\cite{Cor99a,Cor99b,OW_XFe6,OW_CsFe8,Sla02}

Concerning the non-diagonal elements of $\hat{H}_D$, one finds
\begin{equation}
 \langle S|\hat{H}_D|S+1\rangle = \cos\varphi \sin\varphi \sum_i D_i \langle S| \hat{T}^{(2)}_1(i)|S+1\rangle,\nonumber
\end{equation}
where $D_i$ is the projection of the single-ion anisotropy $\mathbf{D}_i$ on the uniaxial axis, and
$\hat{T}^{(2)}_1(i)$ the irreducible tensor operator (ITO) of rank 2 related to the $i$th spin
center.\cite{Mes85,Ben90,Boc99} Accordingly, the angle dependence of the mixing parameter follows
\begin{equation}
\label{eqn:delta}
 \Delta(\varphi) \propto \cos\varphi\sin\varphi,
\end{equation}
so that the level mixing induced by $\hat{H}_D$ is zero for parallel and perpendicular fields, $\varphi$ =
0$^\circ$ and 90$^\circ$, respectively, and maximal for $\varphi$ = 45$^\circ$.

The above first-order results are often a very good approximation. In systems, however, with a large magnetic
anisotropy, such as the CsFe$_8$ wheel,\cite{OW_CsFe8,OW_CsFe8_NVT_INS2} the effects of higher-order
contributions become non-negligible and would have to be considered for a fully quantitative description. For
instance, in second-order an additional quadratic field term $\propto b^2$ would arise. But even then, the
first-order results are useful for qualitative or semi-quantitative considerations.

The inclusion of further magnetic anisotropic terms via $\hat{H}'_1$ produces additional contributions to
$\varepsilon_1$, $\varepsilon_2$, and $\Delta$, i.e., adds the respective matrix elements of $\hat{H}'_1$ to
Eqs.~(\ref{eqn:eps}) and (\ref{eqn:delta}). One term of potential importance is the Dzyaloshinsky-Moriya
interaction $\hat{H}_{DM}$.\cite{Cin02,OW_FIMEI} For this term the diagonal contributions are zero, so that
$\varepsilon_1$ and $\varepsilon_2$ [as well as $b_0(\varphi)$] are not affected. For the non-diagonal
element one finds
\begin{eqnarray}
 \langle S|\hat{H}_{DM}|S+1\rangle &=& \sum_{i \neq j} (i \cos\varphi d_{ijx} + d_{ijy} \cr && -
i \sin\varphi d_{ijz}) \langle S| \hat{T}^{(1)}_1(ij)|S+1\rangle,\nonumber
\end{eqnarray}
where $\hat{T}^{(1)}_1(ij)$ is an ITO of rank 1.\cite{Mes85,Ben90,Boc99} The contribution of $\hat{H}_{DM}$
to the angle dependence of the mixing parameter $\Delta$ thus is quite complicated in general. For a planar
uniaxial cluster with sufficiently high symmetry, such as molecular wheels or grids, only $d_{ijz}$ is
non-zero and $\Delta$ will have a component which varies as $\sin\varphi$.\cite{Cin02}

%

\section{Thermodynamics}

\begin{figure}
\includegraphics[scale=1]{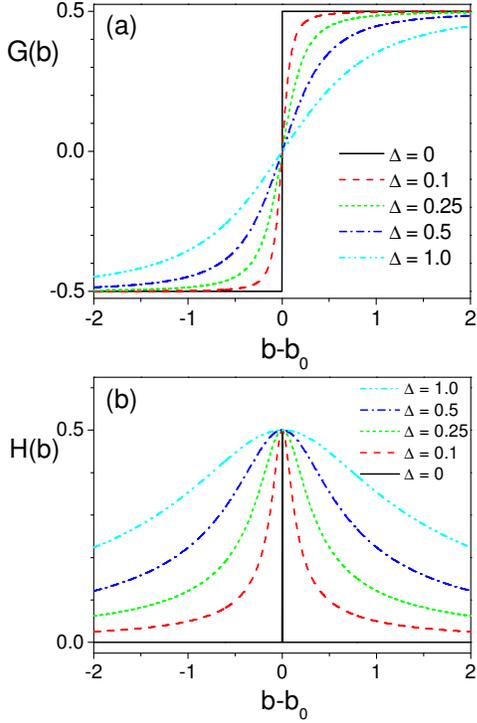}
\caption{\label{fig:g_h_vs_b}(Color online) Field dependence of the functions (a) $G(b,\varphi)$ and (b)
$H(b,\varphi)$ discussed in the text for various values of the level-mixing parameter $\Delta$.}
\end{figure}

The thermodynamic properties for magnetic fields close to a LC can be calculated analytically, keeping in
mind Eq.~(\ref{eqn:eps}) and $\Delta = \Delta(\varphi)$. Since a system with uniaxial magnetic anisotropy is
considered, the free energy $F$ is a function of the temperature $T$, of the magnitude of the magnetic field
$B$ (or $b$), and of its orientation $\varphi$: $F \equiv F(T,b,\varphi)$. In this section, the magnetic
specific heat, the magnetization, and the magnetic torque is calculated. For the calculation it is noted that
$\varepsilon_2-\varepsilon_1 = b_0(\varphi) -b$. It is also convenient to introduce the two functions
\begin{subequations}
\label{eqn:GH}
\begin{eqnarray}
  G(b,\varphi) &=& \frac{b-b_0(\varphi)}{ 2 \sqrt{   [b-b_0(\varphi)]^2 + \Delta^2(\varphi) } },\\
  H(b,\varphi) &=& \frac{\Delta(\varphi)}{ 2 \sqrt{   [b-b_0(\varphi)]^2 + \Delta^2(\varphi) } }.
\end{eqnarray}
\end{subequations}
As function of field, $G(b,\varphi)$ describes a step from -1/2 to 1/2 centered at the LC field $b_0$, which
is broadened by the level mixing due to $\Delta$, as shown in Fig.~\ref{fig:g_h_vs_b}(a). The field
derivative $dG/db$ exhibits a peak at the LC with a height of $1/(2|\Delta|)$ and a
full-width-at-half-maximum (FWHM) of $2 \sqrt{2^{2/3}-1}|\Delta| = 1.124 |\Delta|$.\cite{FWHM} $H(b,\varphi)$
is essentially the square-root of a Lorentz function and accordingly describes a peak at the LC field $b_0$
with a height of $\pm 1/2$, depending on the sign of $\Delta$, and a FWHM of $2\sqrt3 |\Delta| = 3.464
|\Delta|$,\cite{FWHM} see Fig.~\ref{fig:g_h_vs_b}(b). Both $G$ and $H$ can be expressed in terms of the
reduced variable $x = (b-b_0)/|\Delta|$: $G(x) = x H(x)$, $H(x) = (2\sqrt{x^2+1})^{-1}$, which makes the
dependencies on $b_0$ and $\Delta$ transparent.

The magnetic specific heat is given by $c_M = - T \partial^2 F /
\partial T^2$, which with Eq.~(\ref{eqn:Fsb}) yields
\begin{eqnarray}
  c_M(b,\varphi) &=&  \frac{ E_2^2}{k_B T^2\cosh^2(\beta E_2)}
  \cr
  &=& k_B \left(\frac{2 E_2}{k_B T}\right)^2 \frac{\exp(\beta 2 E_2)}{[1+\exp(\beta 2 E_2)]^2}.
\end{eqnarray}
Since $2E_2$ is the energy gap between the two levels, the specific heat exhibits a Schottky anomaly as
function of temperature (with a maximum of height $0.4392 k_B$ at $k_B T = 0.4168 |2E_2|$), and a double-peak
feature as function of field, as was noted and discussed before by several authors, to whose works we hence
refer to.\cite{Aff02,Efr02,Aff03,Efr06,Eva06}

\begin{figure}
\includegraphics[scale=1]{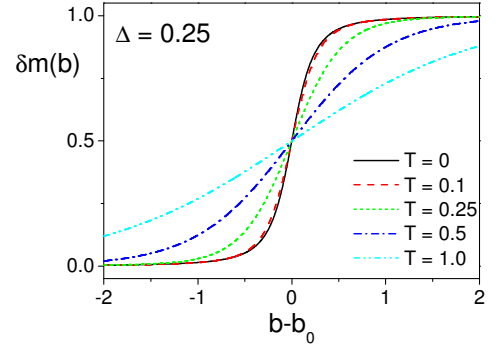}
\caption{\label{fig:dm_vs_b}(Color online) Field dependence of the magnetization change $\delta m(b,\varphi)$
for various temperatures. The level-mixing parameter was set to $\Delta = 0.25$.}
\end{figure}

The magnetization is given by $m = - \partial F / \partial b$ (in units of $g \mu_B$) and calculated to
\begin{equation}
  \label{eqn:m}
  m(b,\varphi) = S + \frac{1}{2} + G(b,\varphi) \tanh(\beta E_2).
\end{equation}
It is convenient to separate the magnetization as $m = m_\ll + \delta m$, where $m_\ll = S$ is the
zero-temperature magnetization at fields well below the LC field, $b \ll b_0$. $\delta m$ is the change in
the magnetization upon passing the LC coming from fields below $b_0$:
\begin{eqnarray}
  \label{eqn:dm}
  \delta m(b,\varphi)
  &=& \frac{1}{2} +  G(b,\varphi) \tanh(\beta E_2) \cr
  &=& \left[\frac{1}{2} + G(b,\varphi)\right] - \left[1-\tanh(\beta E_2)\right] G(b,\varphi).\cr&&
\end{eqnarray}
As function of field, the magnetization exhibits a step at the LC, where it changes from $S$ to $S+1$ (or by
one unit $g \mu_B$). Figure~\ref{fig:dm_vs_b} shows $\delta m(b)$ at various temperatures for $\Delta$ =
0.25. The magnetization can be expressed in terms of the reduced variables $x = (b-b_0)/|\Delta|$ and $t =
k_BT/|\Delta|$ [$\beta E_2 = \sqrt{x^2+1}/(2t)$]. In these units, the variation of the behavior with $b_0$
and $\Delta$ becomes most apparent.

At low temperatures, $k_BT \ll |\Delta|$, the magnetization change is given by $1/2 + G$, and is independent
on temperature. At high temperatures, $k_BT \gg |\Delta|$, it reduces to $1/2 + \tanh[(b-b_0)/(2k_BT)]/2$ as
for a true LC, i.e., the effects of level mixing become indiscernible in this temperature regime.

The magnetization step is broadened by the effects of temperature and non-zero level mixing $\Delta$. The
broadening may be analyzed by considering the width of the peak in the field derivative $dm/db$. At low
temperatures, $k_BT \ll |\Delta|$, the peak is characterized by a temperature-independent FWHM of $1.124
|\Delta|$, and at high temperatures, $k_BT \gg |\Delta|$, by a linearly increasing FWHM of $4 \ln(\sqrt2 + 1)
k_B T = 3.525 k_B T$.\cite{Cor99b,Cin02,FWHM} With temperature, also the shape of the peak changes, so that a
careful analysis of the shape of experimental $dm/dB$ curves in principle provides additional information on
the level mixing.

The magnetic torque is given by $\tau = \partial F / \partial \varphi$,\cite{tau} and calculated
to\cite{sign}
\begin{eqnarray}
  \label{eqn:tau}
  \tau(b,\varphi)
  &=&
  \frac{1}{2} \left( \frac{\partial \Delta_1}{\partial \varphi} + \frac{\partial \Delta_2}{\partial
  \varphi}\right) + \tanh(\beta E_2)
  \cr && \times
   \left[ G(b,\varphi) \frac{\partial b_0}{\partial \varphi}
  + H(b,\varphi) \frac{\partial \Delta}{\partial \varphi} \right] .
\end{eqnarray}
This equation can be transformed into a more useful form by considering i) $(\Delta_1 + \Delta_2 )/ 2 =
\Delta_1 + b_0/2$ and Eq.~(\ref{eqn:dm}), and by separating ii) the torque into $\tau = \tau_\ll + \delta
\tau$, where $\tau_\ll = \partial \Delta_1 / \partial \varphi$ is the torque for fields well below the LC
field, and $\delta \tau$ the change in the torque upon passing the LC. One finds\cite{sign}
\begin{eqnarray}
  \label{eqn:dtau}
  \delta \tau(b,\varphi)
  &=& \delta m(b,\varphi) \frac{\partial b_0}{\partial \varphi}
  + H(b,\varphi) \frac{\partial \Delta}{\partial \varphi} \tanh(\beta E_2) \cr
  &=& \delta m(b,\varphi) \frac{\partial b_0}{\partial \varphi}
  + H(b,\varphi) \frac{\partial \Delta}{\partial \varphi} \cr
  &&- [1-\tanh(\beta E_2)] H(b,\varphi)  \frac{\partial \Delta}{\partial \varphi}.
\end{eqnarray}

Interestingly, the torque consists of two contributions, $\delta \tau = \delta \tau_1 + \delta \tau_2$, where
$\delta \tau_1$ refers to the first term on the r.h.s. of Eq.~(\ref{eqn:dtau}), and $\delta \tau_2$ to the
remaining terms. The first contribution, $\delta \tau_1 $, is proportional to the magnetization change
$\delta m$. It accordingly also produces a step as function of field in the torque signal, which is broadened
by a non-zero level mixing and temperature. The second contribution, $\delta \tau_2$, as it is proportional
to $H(b,\varphi)$, produces a peak in the torque signal centered at the LC field. The field dependence of
$\delta \tau_2$ is depicted in Fig.~\ref{fig:dtau2_vs_b} for different temperatures and several values of
$\Delta$. At low temperatures, $k_BT \ll |\Delta|$, the peak follows $H(b,\varphi)$ and is accordingly
characterized by a temperature-independent height of $(1/2) \partial \Delta /  \partial \varphi$ and FWHM of
$3.646 |\Delta|$.\cite{FWHM} With increasing temperature the peak smears out to disappear at temperatures
$k_BT \gg |\Delta|$. As with the magnetization, the torque is expressible in terms of the reduced variables
$x$ and $t$, clarifying the trends with $\Delta$ and $T$ visible in Fig.~\ref{fig:dtau2_vs_b}. Several points
shall be noted:

\begin{figure}
\includegraphics[scale=1]{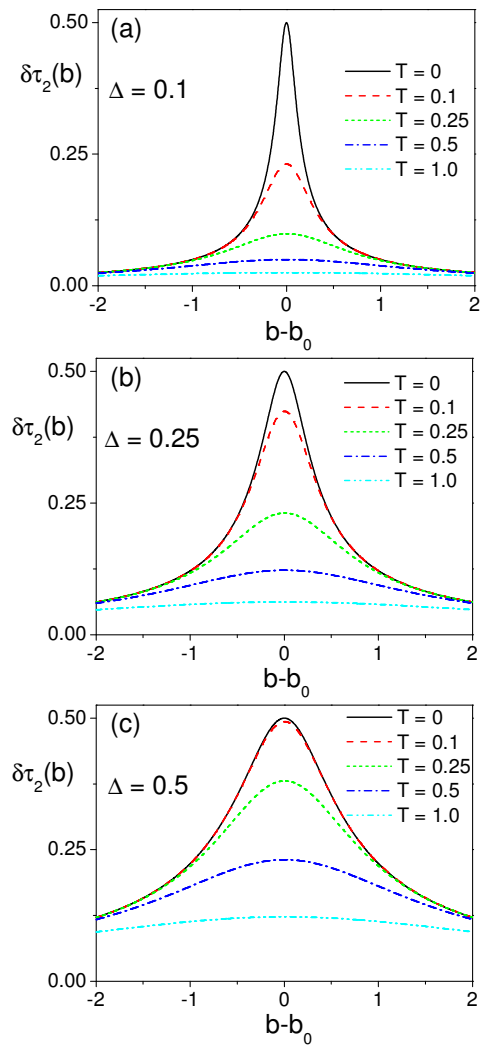}
\caption{\label{fig:dtau2_vs_b}(Color online) Field dependence of the torque contribution $\delta
\tau_2(b,\varphi)$ at several temperatures for the level-mixing parameters (a) $\Delta = 0.1$, (b) $\Delta =
0.25$, and (c) $\Delta = 0.5$.}
\end{figure}

1) For the special case of zero level mixing ($\Delta = 0$), most of the results have been obtained before,
in particular that both the field-dependent magnetization and torque curves exhibit steps at the LC field. In
fact, from Eq.~(\ref{eqn:dtau}) one finds $\delta \tau(b,\varphi) \propto \delta m(b,\varphi)$, since $\delta
\tau_2(b,\varphi) = 0$ for $\Delta = 0$. In reality, however, $\Delta$ never will be exactly zero, but as
long as it is sufficiently small, the torque will be characterized by steps [small means $\partial \Delta /
\partial \varphi \ll \partial b_0 / \partial \varphi$, which is obeyed if $\Delta(\varphi)$ is either
small in magnitude or almost constant as function of angle]. Vice versa, whenever $\tau(b,\varphi)$ is
observed to be mostly step-like, one can conclude that (i) $\Delta$ is very small and (ii) $\delta
\tau(b,\varphi) \propto \delta m(b,\varphi)$. This situation has been observed for instance in many AFM
wheels.\cite{Taf94,Gat94,Cor99a,Cor99b,OW_XFe6,OW_CsFe8,Cin02,Sla02,Aff03}

2) If level mixing, however, is significant, then the magnetization and torque curves in general will be very
different, since then the peak-like contribution $\delta \tau_2(b,\varphi)$ significantly adds to the torque
signal. A similar contribution is not present in the magnetization, and it is thus a unique feature of
torque. This contribution is rooted in the fact that the mixing parameter $\Delta$ does not depend on the
magnetic field but on the angle $\varphi$, and is hence directly probed by torque. If level mixing is
actually very strong, i.e., $\partial \Delta / \partial \varphi \gg \partial b_0 / \partial \varphi$, then
the torque curve will be characterized by peaks at the LCs. At nonzero temperature and/or strong level mixing
the peaks of several nearby LCs may superimpose to yield an oscillatory field dependence of the
torque.\cite{OW_CCR} This situation has been observed for the Mn-[3~$\times$~3] grid and the modified wheels
Cr$_7$Ni and Cr$_7$Zn.\cite{OW_QMO,Car05}

3) Interestingly, if one writes the ground-state wavefunction as
\begin{equation}
  \label{eqn:gs}
 |0\rangle = c |S\rangle + s |S+1\rangle,
\end{equation}
with $c^2 + s^2 = 1$, one finds
\begin{subequations}
  \label{eqn:cs}
\begin{eqnarray}
   \label{eqn:csa}
 s^2 &=& 1/2 + G(b,\varphi),\\
 cs &=& H(b,\varphi).
\end{eqnarray}
\end{subequations}
Hence, at zero temperature, $\delta m(b,\varphi)$ and $\delta \tau_1(b,\varphi)$ directly reflect $s^2$,
while $\delta \tau_2(b,\varphi)$ reflects the product $c s$.

4) From the above points it is clear that the peaks in the torque curves due to $\delta \tau_2(b,\varphi)$
are a direct signature of level mixing, as pointed out previously in Ref.~\onlinecite{Car03b}. In this work
it has been also shown that the peaks in the torque are related to the enhancement of the quantum
fluctuations of the total spin $\hat{\textbf{S}}$ by a level mixing. In fact, a derivation has been given
which yielded $\tau \propto 2B \Delta S_z$, where $(\Delta S_z)^2 = \langle \hat{S}_z^2\rangle - \langle
\hat{S}_z\rangle^2$ denotes the quantum fluctuations in $\hat{S}_z$.\cite{Car03b} This relation cannot be
correct, or complete, since for zero level mixing it would predict $\tau = 0$, in contrast to the
theoretically expected and experimentally observed step-like torque curves. In view of the above findings,
however, $2B \Delta S_z$ can be associated with the contribution $\delta \tau_2$, as both are related to
$cs$.

5) The generic behavior of the angle dependence of the LC field is $b_0(\varphi) \propto (\cos^2 \varphi -
1/3) + const$, so that $\partial b_0 / \partial \varphi \propto \sin \varphi \cos \varphi$. Thus, the
step-like part of the torque will follow $\delta \tau_1(b,\varphi) \propto \delta m(b) \sin \varphi \cos
\varphi$, which is the expected angle dependence.

6) In the above treatment an isotropic $g$ factor was assumed. The calculations for anisotropic $g$ factors
are simple but the results are lengthy. As a general trend, an anisotropic $g$ factor produces an additional
contribution to the torque (and only to the torque) which increases linearly with the magnetic field in first
approximation.\cite{OW_Cu3x3}

%

\section{Magneto-elastic Hamiltonian}

Within the spin-Hamiltonian formalism a magneto-elastic (ME) coupling is introduced by allowing the magnetic
parameters (coupling constants, anisotropy parameters, $g$ factors, etc.) to depend on the distortion
coordinates $\textbf{Q}_k = \textbf{u}_k - \textbf{u}_{k,0}$ of the atoms in the
lattice.\cite{Abr70,Cro79,Pol95,Bec02,Spa04,San05,Elh05} $\textbf{u}_{k,0}$ is the equilibrium and
$\textbf{u}_k$ the distorted position vector of the $k$th atom; the $\textbf{Q}_k$ are usually normal
coordinates. For the exchange coupling constants, for instance, hold
$J_{ij}(\textbf{Q}_1,\textbf{Q}_2,\ldots) = J_{ij} + \delta J_{ij}(\textbf{Q}_1,\textbf{Q}_2,\ldots)$.
Assuming small deviations, a Taylor expansion yields $\delta J_{ij}(\textbf{Q}_1,\textbf{Q}_2,\ldots) \approx
\sum_k {\bf \alpha}_{ij,k} \cdot \textbf{u}_k$, with the ME coupling constants ${\bf \alpha}_{ij,k} \equiv
\partial J_{ij}/\partial \textbf{u}_k$. In the following, it is assumed that only one distortion mode, and
hence only one (scalar) distortion coordinate $Q$, is relevant. For the exchange constant one then finds
$J_{ij}(Q) = J_{ij} + \delta J_{ij}(Q)$ with $\delta J_{ij} = \alpha_{ij} Q$. Similar relations hold for the
other magnetic parameters $\mathbf{D}_i$, $\textbf{d}_{ij}$, $g$, and so on.

Inserting the modulated magnetic parameters into the microscopic spin Hamiltonian produces additional,
$Q$-dependent terms which will be denoted collectively as $\hat{H}_Q$. For the terms considered in this work,
one finds
\begin{eqnarray}
 \label{eqn:HQ}
 \hat{H}_Q &=& -\sum_{i \neq j} \delta J_{ij} \hat{\textbf{S}}_i \cdot \hat{\textbf{S}}_{j}
 + \sum_i \hat{\textbf{S}}_i \cdot \delta \mathbf{D}_i \cdot \hat{\textbf{S}}_i
 \cr &&
 + \sum_{i \neq j} \delta \mathbf{d}_{ij} \cdot (\hat{\mathbf{S}}_i \times \hat{\mathbf{S}}_{j}) + \ldots,
\end{eqnarray}
The combined system of the magnetic molecule and the elastic lattice is then described by the spin-phonon
Hamiltonian
\begin{eqnarray}
 \label{eqn:HSP}
 \hat{H}_{sp-ph} = \hat{H} + \hat{T} + \frac{k}{2} Q^2,
\end{eqnarray}
where $\hat{H}_Q$ is included in $\hat{H}$ via $\hat{H}_1$ or $\hat{H}'_1$, respectively. $\hat{T}$ is the
kinetic energy of the phonons and $k$ the elastic constant.

In this work, the phonons are treated as classical oscillators, which is justified if the lattice dynamics is
much slower than the spin dynamics. This situation is known as the adiabatic limit and has been often used to
infer the ground state in spin-phonon systems.\cite{Bec02,Spa04,Elh05} Under this conditions, the ground
state is obtained for zero kinetic energy of the phonons (atoms at rest) and by minimizing the potential
\begin{eqnarray}
 \label{eqn:V}
 V(Q) = E(Q) + \frac{k}{2} Q^2
\end{eqnarray}
of the total system, where $E(Q)$ is the ground-state energy of the magnetic part $\hat{H}$ (which now
depends on $Q$ due to the inclusion of $\hat{H}_Q$). One may approximate $E(Q) = -a Q^\kappa$, with some
positive constants $a$ and $\kappa$. A stable minimum is obtained for $\kappa < 2$ at $Q_0 =
\sqrt[2-\kappa]{\kappa a/k} > 0$. This situation corresponds to unconditional instability. For $\kappa = 2$
conditional instability is realized, where $V(Q) < 0$ is obtained for any $Q > 0$ if $a > k/2$. For $\kappa >
2$ the only solution is $Q_0 = 0$, i.e., no distortion is obtained.

Within the framework of the TLH approximation, $E(Q) \equiv E_{-}(Q)$ ($E_{-}$ depends on $Q$ via
$\varepsilon_1$, $\varepsilon_2$, and $\Delta$). Apparently, the ME coupling enters in two ways, namely via
the diagonal matrix elements $\langle S|\hat{H}_Q|S\rangle$, and via the non-diagonal matrix element $\langle
S|\hat{H}_Q|S+1\rangle$. The first mechanism shall be called diagonal, the second non-diagonal ME coupling.
These two coupling modes have to be carefully distinguished.

Before proceeding further, a point concerning the Heisenberg-exchange part shall be clarified. Within the
context of the present work it sets the largest energy scale. In several models treated in the literature it
is actually the only term in the spin Hamiltonian.\cite{Rak04,Pop04,Spa04,Elh05,Pop05,Hon06} Two cases have
to be distinguished, namely those where the Heisenberg exchange results in (i) a degenerate or (ii)
non-degenerate ground state. Case (i) is encountered, for instance, in spin triangles or
tetrahedra.\cite{Rak04,Hon06} The description bears some analogy to the Jahn-Teller effect (JTE), and the
phenomenon hence has been denoted frequently as the spin-JTE. Such situations have been analyzed
theoretically several times in the literature,\cite{Rak04,Pop04,Pop05,Hon06} but to the best of our knowledge
no experimental evidence has been reported to date. Case (ii) is clearly distinct, and is the one of interest
in the present work. Here, an unconditional lattice instability is obtained in zero field for gapless spin
systems, such as a spin-1/2 chain (spin-Peierls effect).\cite{Cro79} For gapped systems, such as the finite
spin clusters considered here, a spontaneous instability typically does not occur. With appropriate magnetic
fields a degenerate ground state can be created via the LC mechanism, but for a purely isotropic model a
spontaneous instability nevertheless does not occur: The degeneracy at the LC points cannot be lifted by
whatever distortion since the two involved levels are eigenstates of $\hat{\textbf{S}}$ with different values
of $S$. They hence cannot be mixed by the Heisenberg exchange. This already indicates the crucial role of a
magnetic anisotropy in $\hat{H}_1$ for a spontaneous instability at a LC in case-(ii) systems.

The analysis of the diagonal ME coupling requires the evaluation of $\langle S|\hat{H}_Q|S\rangle$. There is
no general reason why $\langle S|\hat{H}_Q|S\rangle$ should be zero, but a non-zero value would lead to
somewhat unphysical situations. The effect of a diagonal ME coupling would be a shift of the LC fields by a
distortion, $b_0 = b_0(Q)$. A change of $b_0$, however, corresponds to a change of the zero-field splitting,
see Eq.~(\ref{eqn:eps}). For the exchange coupling, e.g., this would imply a change of the average coupling
strength, i.e., of the overall magnetic energy scale. This is not what one normally expects. For instance,
for a ring-like model the generic distortion mode would be a dimerization, upon which the average $\sum_i
J_{i,i+1}$ is not altered. Hence $\sum_i \delta J_{i,i+1} = 0$, implying a zero diagonal element. Similar
arguments can be put forward for the other magnetic terms ($\hat{H}_{DM}$ gives zero anyhow). Thus, for
typical physical models the diagonal ME coupling is zero in first-order. Second-order terms, however, are
always expected to be non-zero, and one accordingly concludes that
\begin{equation}
  \langle S|\hat{H}_Q|S\rangle \propto Q^2,
\end{equation}
corresponding to conditional instability. These considerations are in agreement with the findings of
Refs.~[\onlinecite{Spa04,Elh05}]. For a gapped quantum spin system one expects the generic behavior $E(Q)
\propto Q^2$.

The situation is different for the non-diagonal ME coupling $\langle S|\hat{H}_Q|S+1\rangle$. For the
Heisenberg contribution in $\hat{H}_Q$ the matrix element is strictly zero, but for the anisotropic terms it
will be non-zero in general. This is demonstrated by the simplified example of a regular AFM spin ring, for
which a Dzyaloshinsky-Moriya interaction emerges upon a dimerization, $d_{i,i+1,z}(Q) = (-1)^i \alpha Q$
($d_{i,i+1,x}=d_{i,i+1,y}=0$). Then $\langle S|\hat{H}_Q|S+1\rangle \propto \sum_i (-1)^i \alpha Q \langle
S|\hat{T}^{(1)}_1(i,i+1)|S+1\rangle \propto Q$, since $\langle S|\hat{T}^{(1)}_1(i,i+1)|S+1\rangle$ =
$(-1)^i$ $\langle S|\hat{T}^{(1)}_1(1)|S+1\rangle$. The point is, in a qualitative language, that the
diagonal elements essentially probe the \emph{average} of the modulation (in our example $\sum_i \delta
d_{i,i+1,z} = 0$), while the non-diagonal matrix element is sensitive to the \emph{span} of the modulation
[in our example $\sum_i (-1)^i \delta d_{i,i+1,z} \neq 0$]. Hence, in general,
\begin{equation}
  \langle S|\hat{H}_Q|S+1\rangle \propto Q + const,
\end{equation}
which allows for unconditional instability.

The above considerations imply the following TLH to describe the ME effects at a LC:
\begin{eqnarray}
 \label{eqn:Hsp2}
 &\hat{H}_{SP} = \left(\begin{array}{cc} \varepsilon_S(b,\varphi) & \Delta(\varphi,Q)/2 \cr
 \Delta(\varphi,Q)/2 & \varepsilon_{S+1}(b,\varphi) \end{array} \right) +  \frac{k}{2} Q^2,&
 \cr&&\cr&&\cr
 &\Delta(\varphi,Q) = \Delta(\varphi,0) + \alpha(\varphi) Q.&
\end{eqnarray}
The essential elements of this model are that (i) the diagonal elements do not depend on the distortion $Q$,
(ii) the non-diagonal elements do not depend on the magnetic field $b$, and (iii) $\Delta(Q)$ varies linearly
with $Q$.

The model explicitly allows for the possibility of a non-zero non-diagonal matrix element even for the
non-distorted molecule, i.e, for $\Delta(0) \neq 0$.\cite{distortion} It is easy to show that the system
exhibits an unconditional instability for any value of $\Delta(0)$. However, the most interesting situation
arises for sufficiently small $\Delta(0)$ (what small means will be made precise later), and this case is
also the one of relevance for CsFe$_8$.\cite{DMinCsFe8} The behavior for small $\Delta(0)$ is not essentially
different from that for zero $\Delta(0)$, and in the following $\Delta(0) = 0$ is hence always assumed. The
case of $\Delta(0) \neq 0$ will be considered at the end of the next section. The other approximations and
limitations leading to $\hat{H}_{SP}$ have been carefully discussed in the above.

Formally, $\hat{H}_{SP}$ is equivalent to the standard TLH discussed in the JTE. Indeed, the problems are in
many respects similar, and much of the insights and results developed for the JTE can be directly carried
over to the current problem (key words to mention are cooperative-JTE, pseudo-JTE, etc.).\cite{GehBarBer} The
physical difference, however, is that the JTE is related to a degeneracy in the electronic system, while here
it originates from a degeneracy in the spin system. The effect described by $\hat{H}_{SP}$ hence might be
called a spin-JTE. This notation is not quite satisfying, since the underlying mechanism is different from
the spin-JTE for Heisenberg spin systems with a degenerate spin ground state, which was mentioned before. The
effect discussed here therefore is tentatively called field-induced spin-JTE (FISJTE).

An interesting feature, as compared to the electronic and spin-JTE, is the possibility to continuously tune
through the instability via an applied magnetic field. At the LC, $\varepsilon_{S+1}-\varepsilon_{S}$ is
zero, but can be adjusted to any value by moving away from the LC point, realizing a situation similar to the
pseudo-JTE. In the next section, the behavior of the system as function of magnetic field is considered in
detail.

\section{Field-induced magneto-elastic instability}

\begin{figure}
\includegraphics[scale=1]{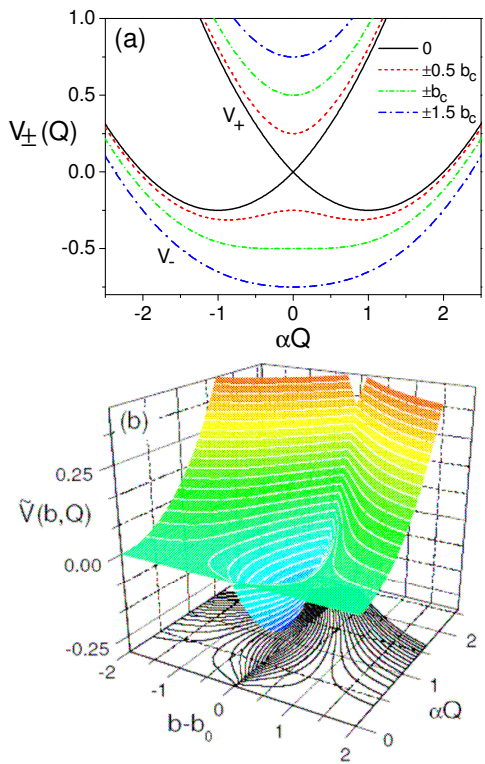}
\caption{\label{fig:v_vs_b_q} (Color online) (a) Distortion dependence of the two PES $V_\pm(b,Q)$ for
several fields $b-b_0$ and (b) field and distortion dependence of the potential $\tilde{V}(b,Q) = V_-(b,Q) -
E_-(b,0)$ for positive values of $Q$ ($b_c = \Delta_0^{max} = 1$, $k = 1/2$).}
\end{figure}

For the discussion of the field-induced ME instability it is convenient to first describe the potential
energy surfaces (PES) produced by $\hat{H}_{SP}$. Two PES are obtained, $V(b,Q)_\pm = E_\pm(b,Q) + (k/2) Q^2$
(where of course $V_- \equiv V$). The dependence on $Q$ is shown in Fig.~\ref{fig:v_vs_b_q}(a) for several
magnetic fields. At the LC, $b-b_0 = 0$, the PES are degenerate at $Q = 0$ and exhibit two minima at
$Q_0^{max} = \alpha /(2k)$. With increasing distance from the LC field, i.e., increasing $|b-b_0|$, a gap
opens between the PES due to the Zeeman splitting, and the two minima accordingly shift to lower $Q$ values.
For fields below $b_0 - b_c$ or above $b_0 + b_c$, where $b_c = \alpha^2 /(2k)$, the minima disappear
completely (in the context of the electronic JTE, $b_c$ is the JT energy).

The spontaneous distortion in the ground state is determined by the minima in the PES, $\partial V /
\partial Q = 0$. It is more convenient, however, to discuss $\tilde{V}(Q) = V(Q) - E_-(Q=0)$, which directly
measures the gain in potential energy due to a distortion. $\tilde{V}$ is determined to
\begin{equation}
 \label{eqn:V2}
 \tilde{V}(b,Q) = - \frac{1}{2} \sqrt{(b-b_0)^2 + \Delta(Q)^2}  + \frac{1}{2} |b-b_0|  + \frac{k}{2} Q^2.
\end{equation}
The field and distortion dependence of $\tilde{V}$ is shown in Fig.~\ref{fig:v_vs_b_q}(b) for positive values
of $Q$ (and $k=1/2$). One clearly observes a field range around the LC where $\tilde{V}$ is negative for
non-zero $Q$. At these fields the system gains energy by distorting to a finite distortion $Q$, i.e., a
spontaneous instability occurs.

\begin{figure}
\includegraphics[scale=1]{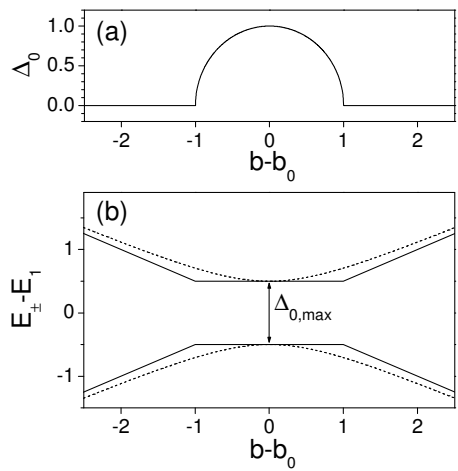}
\caption{\label{fig:de_vs_b} Field dependence of (a) the equilibrium mixing parameter (or order parameter)
$\Delta_0$ and (b) the energies $E_{\pm}-E_1$ of the two levels ($b_c = \Delta_0^{max} = 1$, $k = 1/2$). The
dashed lines in (b) show the energies of an ordinary avoided LC with a mixing parameter $\Delta =
\Delta_0^{max}$.}
\end{figure}

From minimizing $V(Q)$, or $\tilde{V}(Q)$, the condition for the equilibrium distortion $Q_0$, or the
equilibrium mixing $\Delta_0 \equiv \Delta(Q_0) = \alpha Q_0$, respectively, is obtained as
\begin{eqnarray}
 \label{eqn:Vmin}
 (b-b_0)^2 + \Delta_0^2 = \left(\frac{\alpha^2}{2k}\right)^2.
\end{eqnarray}
This is of the form $x^2 + y^2 = r^2$, hence the mixing $\Delta_0$ (or $Q_0$, either one can be used as the
order parameter) describes a semi-circle with radius $b_c$ as function of the relative magnetic field
$b-b_0$, as shown in Fig.~\ref{fig:de_vs_b}(a). For fields below $b_0 - b_c$ and above $b_0 + b_c$, one finds
$\Delta_0 = 0$ and the system is undistorted. For fields in the range $[b_0 - b_c,b_0 + b_c]$, however,
$\Delta_0$ becomes non-zero signaling the spontaneous instability. At exactly the LC, the mixing and the
distortion are maximal, assuming the values $\Delta_0^{max}$ and $Q_0^{max}$ ($\Delta_0^{max} = \alpha
Q_0^{max} = b_c = \alpha^2 /(2k)$).

Concerning the ground-state wavefunction $|0\rangle$, Eq.~(\ref{eqn:gs}), the coefficient $s = \langle
S+1|0\rangle$ is calculated to $s^2 = [1+ (b-b_0)/b_c]/2$ by inserting Eq.~(\ref{eqn:Vmin}) into
Eq.~(\ref{eqn:csa}). Thus, in the field range $[b_0 - b_c,b_0 + b_c]$ the parameter $s^2$ increases linearly
from 0 to 1 as function of field (and is 0 and 1, respectively, for fields outside this range).

With Eqs.~(\ref{eqn:E}) and (\ref{eqn:Vmin}), the energies of the two levels are calculated to $E_{\pm} = (
\varepsilon_2 + \varepsilon_1 \pm \Delta_0^{max} )/2$ for fields in $[b_0 - b_c,b_0 + b_c]$. Outside this
field range, the energies exactly correspond to that of a true LC. The field-dependence of $E_{\pm}-E_1$ is
shown in Fig.~\ref{fig:de_vs_b}(b). Thus, the distortion is such as to maintain a minimum gap of value
$\Delta_0^{max}$ between the energy levels. For comparison also the energy diagram of a conventional avoided
LC with a mixing $\Delta_0^{max}$ is shown, in order to emphasize the distinctly different behavior of the
present model. In an avoided LC, the mixing is field independent. In the present model, in contrast, the
mixing is zero for fields outside the range $[b_0 - b_c,b_0 + b_c]$, and non-zero within it due to the
instability, see Fig.~\ref{fig:de_vs_b}(a) (it is recalled that the mixing is directly related to the
spontaneous distortion via $\Delta_0 = \alpha Q_0$).

\begin{figure}
\includegraphics[scale=1]{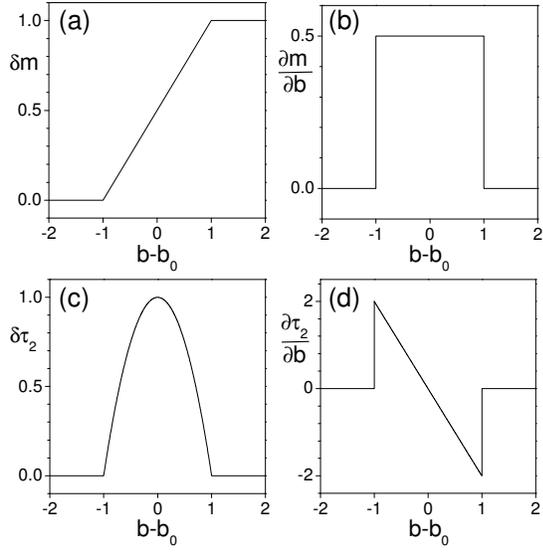}
\caption{\label{fig:mtau_vs_b} Field dependence of (a) the magnetization change $\delta m$, (b) the
field-derivative of $\delta m$, (c) the torque contribution $\delta \tau_2$, and (d) the field derivative of
the torque contribution $\delta \tau_2$ ($b_c = \Delta_0^{max} = 1$, $k = 1/2$).}
\end{figure}

The model permits an analytical calculation of the thermodynamic functions at zero temperature (or
sufficiently low temperatures, $k_B T \ll \Delta_0^{max}$). The specific heat is zero, since the energy
levels never come closer than $\Delta_0^{max}$. Concerning the calculation of the magnetization and torque,
the two functions $G$ and $H$, Eq.~(\ref{eqn:GH}), simplify to $G = (b-b_0)/(2b_c)$ and $H =
\Delta_0/(2\Delta_0^{max})$ for fields in the range $[b_0 - b_c,b_0 + b_c]$. Outside this range they become
$G = \pm 1/2$ and $H=0$. Inserting this in Eqs.~(\ref{eqn:dm}) and (\ref{eqn:dtau}) (and considering $T
\rightarrow 0$), one obtains\cite{sign}
\begin{subequations}
\begin{eqnarray}
 \label{eqn:fff}
 \delta m(b,\varphi) &=& \frac{1}{2} \left[ 1+ \frac{b-b_0}{b_c}\right],
 \\\cr
 \delta \tau(b,\varphi) &=& \frac{1}{2} \left[ 1+ \frac{b-b_0}{b_c}\right] \frac{\partial b_0}{\partial \varphi}
 + \frac{1}{2} \frac{\Delta_0}{\Delta_0^{max}} \frac{\partial \Delta_0}{\partial \varphi}.\qquad
\end{eqnarray}
\end{subequations}
As function of field, the magnetization change $\delta m$ increases linearly from $\delta m = 0$ at $b = b_0
- b_c$ to $\delta m = 1$ at $b = b_0 + b_c$, as shown in Fig.~\ref{fig:mtau_vs_b}(a).
Figure~\ref{fig:mtau_vs_b}(b) presents the field derivative of the magnetization, $\partial m /
\partial b$.

As emphasized in Sec.~IV, the torque consists of two contributions, $\delta \tau = \delta \tau_1 + \delta
\tau_2$. The first contribution is simply proportional to the magnetization change, $\delta \tau_1 = \delta m
(\partial b_0 / \partial \varphi)$, see Eq.~(\ref{eqn:dtau}), and hence also increases linearly in the field
range $[b_0 - b_c,b_0 + b_c]$. The field dependence of the second contribution needs more consideration. As
will be discussed in a moment, the term $\partial \Delta_0 /
\partial \varphi$ should be interpreted as $(\partial \alpha / \partial \varphi)Q_0$. With respect to the
field dependence it is hence proportional to $\Delta_0$. This yields $\delta \tau_2 \propto \Delta_0^2$. The
field dependence of $\delta \tau_2$ and its field derivative are shown in Figs.~\ref{fig:mtau_vs_b}(c) and
\ref{fig:mtau_vs_b}(d), respectively. $\delta \tau_2$ displays a dome-shaped behavior in the field range
$[b_0 - b_c, b_0 + b_c]$. Both types of torque curves, "linear slope" and "dome shaped", were observed in
experiments on the AFM wheel CsFe$_8$.\cite{OW_FIMEI}

In general, since $\delta \tau = \delta \tau_1 + \delta \tau_2$, the torque is the sum of a linear slope and
a dome-shaped contribution. The respective weights are governed by the factors $\partial b_0 /
\partial \varphi$ and $(\partial \alpha / \partial \varphi)/\alpha$, or the ratio of them, respectively. The
actually observed torque profiles hence will depend on the angle $\varphi$ in general, as well as on the
particular system under consideration. A careful investigation of the angle dependence of the torque profiles
thus can provide information on the relevant ME coupling terms in $\hat{H}_Q$.

At this point an apparent contradiction shall be addressed: In the field range $[b_0 - b_c, b_0 + b_c]$ the
energies $E_\pm$ as well as the magnetization exhibit a linear slope, in conflict with the relation $m = -
\partial E / \partial b$ valid at zero temperature. Also, the above procedure of just inserting
Eq.~(\ref{eqn:Vmin}) into the equations of Sec.~IV seems questionable, since $\Delta_0$ is implicitly field
(and angle) dependent via Eq.~(\ref{eqn:Vmin}). This violates the assumption $\partial \Delta /\partial b =
0$ in the derivations of Sec.~IV. However, strictly spoken the PES $V_\pm$ should be used in the
calculations, and not $E_\pm$, since the combined system of spins and lattice vibrations is considered. This
would add terms due to $\partial Q_0^2/ \partial b$ and $\partial Q_0^2/ \partial \varphi$ to the
magnetization and torque. Such a more rigorous calculation, however, yields exactly the same results as with
using the equations of Sec.~IV and neglecting the implicit field and angle dependence of $\Delta_0$. That is,
for the thermodynamic functions $\Delta_0 = \alpha Q_0$ can be used in the sense of $\partial Q_0 /\partial b
= \partial Q_0 /\partial \varphi = 0$. The energy spectrum shown in Fig.~\ref{fig:de_vs_b}(b) is not suited
to derive the thermodynamics, but correctly reflects the gap as it would be observed in spectroscopic
measurements.

The behavior for non-zero temperatures may be treated along the lines of, for instance,
Ref.~\onlinecite{Pop04}. With the two PES $V_\pm$, the free energy $F \equiv F(T,b,\varphi,Q)$ becomes
\begin{equation}
F = E_1 + \frac{k}{2}Q^2 - \frac{1}{\beta} \ln\!\!\left[2\cosh\!\!\left(\frac{\beta}{2} \sqrt{ (b-b_0)^2 +
(\alpha Q)^2}\right)\right].
\end{equation}
The equilibrium distortion $Q_0$ is determined by $\partial F /\partial Q = 0$. It is convenient to switch to
the order parameter $\Delta = \alpha Q$, and to express the resulting condition in terms of the reduced
variables $\bar{b} = (b-b_0)/b_c$, $\bar{\Delta}_0 = \Delta_0/\Delta_0^{max}$, and $\bar{T} = T/T_c^{max}$
(where $k_B T_c^{max} = b_c/2$; we recall $\Delta_0^{max} = b_c$). In these units, one obtains
\begin{subequations}
\label{eqn:tanhX}
\begin{eqnarray}
  \tanh(\bar{X}/\bar{T}) =  \bar{X},\\
  \bar{X} \equiv \sqrt{ \bar{b}^2 + \bar{\Delta}_0^2}.
\end{eqnarray}
\end{subequations}
This sort of transcendent equation often appear in mean-field theories, for instance the Weiss
molecular-field theory of ferromagnets.\cite{Ash76} This shows that the above theory actually establishes the
mean-field theory of the FISJTE.

\begin{figure}
\includegraphics[scale=1]{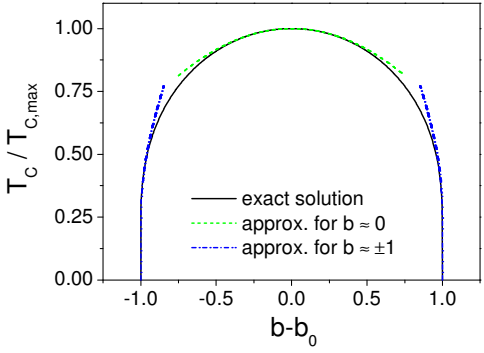}
\caption{\label{fig:tc_vs_b} (Color online) Field dependence of the reduced critical temperature $\bar{T}_c$
(solid line). The dashed (green online) and dash-dotted (blue online) lines represent the approximate
solutions $\bar{b} \approx \sqrt{3(1-\bar{T}_c)}$ for $|\bar{b}| \ll 1$ and $\bar{T}_c \approx 2
\ln[2/(1-|\bar{b}|)]^{-1}$ for $\bar{b} \rightarrow \pm 1$, respectively.}
\end{figure}

At the LC, where $\bar{b} = 0$ and hence $\bar{X} \propto \bar{\Delta}_0$, the graphical solution of
Eq.~(\ref{eqn:tanhX}) proceeds exactly as in textbooks on the Weiss ferromagnet.\cite{Ash76} The critical
temperature, below which a spontaneous distortion occurs, is determined by $\tanh(\bar{X}/\bar{T}_c) \approx
\bar{X}/\bar{T}_c = \bar{X}$, which yields $\bar{T}_c = 1$ or $k_B T_c = k_B T_c^{max} = b_c/2$,
respectively. At the LC, $T_c$ assumes its maximal value, $T_c^{max}$. For fields below or above the LC,
$T_c$ is reduced, in accord with the finding that the distortion is largest at the LC point at zero
temperature [see Fig.~\ref{fig:de_vs_b}(a)].

The zero-temperature behavior is easily rederived. For $T = 0$, Eq.~(\ref{eqn:tanhX}) reduces to the
condition $\bar{b}^2 + \bar{\Delta}_0^2 = 1$, which is equivalent to Eq.~(\ref{eqn:Vmin}).

The critical temperature for an arbitrary field is given by the solution of Eq.~(\ref{eqn:tanhX}) for
$\bar{\Delta}_0 = 0$, i.e., $\tanh(\bar{b}/\bar{T}_c) = \bar{b}$, which yields
\begin{eqnarray}
\label{eqn:Tc}
  \bar{T}_c(\bar{b}) = 2 \bar{b} \ln\left[ \frac{1+\bar{b}}{1-\bar{b}}\right]^{-1}.
\end{eqnarray}
The dependence of $\bar{T}_c$ on $\bar{b}$ (or $b$, respectively) is displayed in Fig.~\ref{fig:tc_vs_b}.
Near the LC point, where $|\bar{b}| \ll 1$, one finds $\bar{b} \approx \sqrt{3(1-\bar{T}_c)}$, corresponding
to a critical exponent of 1/2 as expected for a mean-field theory. Near the upper or lower critical fields,
$\bar{b} \rightarrow \pm 1$, the critical temperature $\bar{T}_c$ goes to zero according to $\bar{T}_c
\approx 2 \ln[2/(1-|\bar{b}|)]^{-1}$. The field dependencies of these approximations are also displayed in
Fig.~\ref{fig:tc_vs_b}.

The analysis so far was based on the assumption of $\Delta(Q) = \alpha Q$ or $\Delta(0) = 0$, respectively.
In real materials, however, $\Delta(Q) = \Delta(0) + \alpha Q$ could be more appropriate. For instance, in
the AFM wheel CsFe$_8$ the molecular C$_4$ symmetry permits next-neighbor Dzyaloshinsky-Moriya interactions,
which would result in $\Delta(0) \neq 0$ [for the specific example of CsFe$_8$, however, experiments indicate
that $\Delta(0)$ is in fact rather small\cite{DMinCsFe8}]. The effect of a non-zero value of $\Delta(0)$ at
zero temperature shall be shortly discussed.

\begin{figure}
\includegraphics[scale=1]{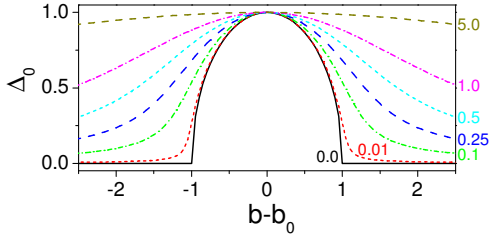}
\caption{\label{fig:d_vs_b_pd0} (Color online) Field dependence of the order parameter $\Delta_0$ for the
indicated values of the ratio $\Delta(0) / \Delta_0^{max}$ ($b_c = \Delta_0^{max} = 1$, $k = 1/2$).}
\end{figure}

The minimization of $V(Q)$ as in the above, but with $\Delta(Q) = \Delta(0) + \alpha Q$, reveals that the
system is unconditionally instable for any magnetic field, and not only in the range $[b_0-bc,b_0+b_c]$ as in
the case of $\Delta(0) = 0$. This is expected since the unconditional instability is not related to
$\Delta(0) = 0$, but to the presence of a non-diagonal ME coupling. An analytic solution is not possible for
general values of $\Delta(0)$. Figure~\ref{fig:d_vs_b_pd0} hence displays the numerically calculated field
dependencies of the order parameter $\Delta_0$ for various values of the ratio $\Delta(0) / \Delta_0^{max}$.
For $\Delta(0) = 0$ of course the previous result is reproduced. Although the system is unconditionally
instable in the whole field range for any value $|\Delta(0)| > 0$, Fig.~\ref{fig:d_vs_b_pd0} reveals that for
small $\Delta(0)$, i.e., $\Delta(0) \ll \Delta^{max}_0$, the distortion outside of the field range
$[b_0-bc,b_0+b_c]$ is very small as compared to the maximum distortion $Q^{max}_0 = \Delta_0^{max}/\alpha$.
The behavior remains very similar to the one obtained for $\Delta(0) = 0$, only the onsets of the distortion
at the fields $b_0-b_c$ and $b_0+b_c$ are less sharp. Hence, the magnetization and torque curves will look
similar as in Fig.~\ref{fig:mtau_vs_b}, but with the sharp features at $b_0-b_c$ and $b_0+b_c$ rounded by a
small $\Delta(0)$.

For large values of $\Delta(0)$, $\Delta(0) \gg \Delta^{max}_0$, the situation changes. In this limit the
minimization of $V(Q)$ yields for the equilibrium distortion the simple result $Q_0 = (\alpha/k) H(b)$, where
in the function $H(b)$, Eq.~(\ref{eqn:GH}b), one has to insert $\Delta(0)$ for the level-mixing parameter
$\Delta$. In terms of the order parameter $\Delta_0$ this corresponds to $\Delta_0 = 2\Delta_0^{max} H(b)$,
which shows that $\Delta_0$ will not exceed $\Delta_0^{max}$, and in particular will remain much smaller than
$\Delta(0)$. Physically $-$ it is now more intuitive to think in terms of a small $\Delta^{max}_0$ or weak ME
coupling, respectively $-$ the coupling of the spin system to the lattice is so weak that a lattice
distortion does not affect the energy of the system much. The distortion traces the (field-dependent) energy
gap, as encoded in $H(b)$, but does not modify it significantly. The magnetization and torque curves thus
essentially show the profiles found in Sec.~IV for zero ME coupling, with a broadening parameter $\Delta =
\Delta(0)$. Accordingly, for $\Delta(0) \gg \Delta^{max}_0$ a ME coupling will not lead to pronounced
effects. This is reasonable, it just means that in order to raise the effects of the field-induced ME
instability to an observable level, the system should exhibit a sufficiently strong coupling to the lattice
and/or sufficiently soft vibration mode.

%

\begin{acknowledgments}
Financial support by EC-RTN-QUEMOLNA, contract n$^\circ$ MRTN-CT-2003-504880, and the Swiss National Science
Foundation is acknowledged.
\end{acknowledgments}

%

%

\begin{references}
\bibitem{SM}
The extension of the results of this work to the case of general values of $S$ for the two levels is obvious.
\bibitem{FWHM}
The given values for the FWHM refer to $b = g \mu_B B$ as axis of abscissae. For a comparison with
experimental results, as function of $B$, the FWHM values should be divided by a factor $g \mu_B$.
\bibitem{sign}
In the previous references [\onlinecite{OW_QMO}] and [\onlinecite{OW_FIMEI}], the sign before the terms
proportional to $\partial\Delta/\partial\varphi$ were given incorrectly.
\bibitem{distortion}
In this work distortion refers to a distortion of the molecule's structure as compared to the structure
without ME coupling. At zero-field or far away from the LC, respectively, the molecule may be "distorted" in
the sense that its structure exhibits deviations from some idealized geometry, such as a perfect cyclic
symmetry in the case of AFM wheels, but these "distortions" are not considered in this work as the result of
a ME coupling. The presented model describes the (additional) field-induced distortions near a LC. It does
not put any restrictions on the possible zero-field (i.e., non-distorted) structures of the molecule, besides
that the molecule is magnetically uniaxial.
\end{references}
\end{document}